\documentclass[preprint,aps, prb, floatfix]{revtex4}
\usepackage{graphicx}

\begin{document}
\title{Linear, third- and fifth-order nonlinear spectroscopy of a charge transfer system coupled to an underdamped vibration}
\author{Arend~G.~Dijkstra$^{(a)}$}
\author{Yoshitaka Tanimura$^{(b)}$}
\affiliation{a) Max Planck Institute for the Structure and Dynamics of Matter, Hamburg, Germany}
\affiliation{b) Department of Chemistry, Kyoto University, Kyoto, Japan}
\begin{abstract}
We study hole, electron and exciton transport in a charge transfer system in the presence of underdamped vibrational motion. We analyze the signature of these processes in the linear and third-, and fifth-order nonlinear electronic spectra. Calculations are performed with a numerically exact hierarchical equations of motion method for an underdamped Brownian oscillator spectral density. 
We find that combining electron, hole and exciton transfer can lead to non-trivial spectra with more structure than with excitonic coupling alone. Traces taken during the waiting time of a two-dimensional spectrum are dominated by vibrational motion and do not reflect the electron, hole, and exciton dynamics directly. We find that the fifth-order nonlinear response is particularly sensitive to the charge transfer process. While third-order 2D spectroscopy detects the correlation between two coherences, fifth-order 2D spectroscopy (2D population spectroscopy) is here designed to detect correlations between the excited states during two different time periods.
\end{abstract}

\date\today
\maketitle

\section{Introduction}

Charge transfer lies at the heart of the function of biological photo-synthetic light harvesting complexes, as well as organic solar cells. An optically created electron hole pair (exciton) must be separated into free charges to enable the function of the system. The reaction center or an interface between two materials is the place where the separation process takes place, followed by a process that utilized the charges. In addition to these materials, charge transfer is also important for the nucleobases in DNA and for model systems.

Charge transfer in these systems is thought to be mediated by vibrations.\cite{Barbara.1992.science.256.975, Delor.2014.science.346.1492} Vibrations can provide the energy fluctuations needed to bring exciton and charge transfer states in resonance, as well as dissipate the excess energy. A model that is often employed to model electron transfer in biological and chemical systems uses a single nuclear reaction coordinate, which is in turn coupled to many other degrees of freedom.\cite{Garg.1985.jcp.83.4491, Yan.1988.jpc.92.4842} The coupling to the environment leads to friction, which can influence the dynamics of the electron transfer process.

Electron transfer can be studied experimentally by nonlinear optical spectropscopy,\cite{Kliner.1992.jacs.114.8323, Walker.1992.jpc.96.3728, Miller.2010.jpca.114.2117, McMahon.1998.bpj.74.2567} and, in particular, two-dimensional optical spectroscopy.\cite{Myers.2010.jpcl.1.2774, Bixner.2012.jcp.136.204503, Romero.2014.natphys.10.676}
Over the past decade, ultrafast two-dimensional spectroscopy\cite{Jonas.2003.arpc.54.425} has been developed as a tool to study energy transfer in complex systems in real time.\cite{Songaila.2013.jpcl.4.3590, Panitchayangkoon.2011.pnas.108.20908} Oscillating signatures were found which were interpreted as a signature of electronic coherence in light-harvesting systems,\cite{Engel.2007.nature.446.782, Collini.2010.nature.463.644} as well as conjugated polymers.\cite{Collini.2009.science.323.369} Recently, it was realized that underdamped vibrations can also explain part or all of the observed oscillating signal.\cite{Tiwari.2013.pnas.110.1203, Hayes.2013.science.430.6139, Halpin.2014.natchem.6.196, Chin.2013.nphys.9.113, Jun.2014.jpcl.5.1386} The interplay of delocalized exciton states with coupling to vibrations greatly influences the energy transport.\cite{Dijkstra.2015.jpcl, Vegte.2015.jpcb.119.1302, Plenio.2008.njp.10.113019, Wu.2010.njp.12.105012, Valleau.2012.jcp.137.224103} Similar effects have been observed in artificial J-aggregates.\cite{Milota.2013.jpca.117.6007} It is natural to ask the question how charge transfer processes are reflected in these spectroscopies, and what information can be obtained from them.

Calculated two-dimensional spectra of excitonic systems coupled to vibrations have been reported.\cite{Egorova.2007.jcp.126.074314, Huynh.2013.jcp.139.104103} In general, they exhibit cross peaks which correspond to the vibrational side peaks in the linear spectrum. It is not clear how this structure in the spectrum can be used to learn something about energy or charge transport. In particular, because underdamped vibrations play an important role in the charge transfer process, on would like to use two-dimensional spectroscopy to study this fundamental process. The ultimate goal is to understand how nature uses electronic and vibronic coherence to enable the function of photosynthetic systems, and how these principles can be used to guide the design of materials for light harvesting applications.

In this paper, we set out to theoretically study two-dimensional infrared spectra of a system where excitons and charge transfer coexist.\cite{Wahadoszamen.2014.ncomm.5.5287, Novoderezhkin.2007.bpj.93.1293, Romero.2012.bpj.103.185} The charge transfer process is mediated by coupling to an underdamped vibration, which, in turn, is coupled to a dissipative bath. We calculate, for the first time, the dynamics of charge separation and two-dimensional optical spectra in the presence of vibrations. Although two-dimensional optical spectra of an electron transfer system were calculated in previous work,\cite{Tanimura.2012.jcp.137.22A550} the properties of spectra in the presence of both charge and exciton transfer, which is relevant to real systems, is still an open problem, which we address here.
Our hierarchical equations of motion approach allows us to incorporate strong coupling to vibrations as well as a proper treatment of system bath coherence, which is important in the modelling of two-dimensional optical spectra. 

The remainder of this paper is organized as follows. In section \ref{sec:model} we present the model used in the calculations. In section \ref{sec:results} we discuss the calculated linear and two-dimensional optical spectra for a system where charge transfer and exciton transport coexist. In section \ref{sec:concl} we present our conclusions.

\section{Model} \label{sec:model}

Because the natural and artificial systems that form the topic of this paper have an enormously complicated structure, it is impossible to treat all degrees of freedom quantum mechanically in a dynamic model. However, in order to understand fundamental concepts such as electronic and vibronic coherence, a quantum mechanical model of the functional part of the system is required. The usual way out of this problem is to model the system of relevant electronic degrees of freedom quantum mechanically, while the vibrational environment is treated as a bath. In our case, the system will be the exciton as well as charge transfer states. While it is possible to model the environment using the laws of classical physics, this is not good enough for our current purpose. The reason is that one of our aims is to assess the role of vibronic coherence, which is a quantum effect that exists as quantum coherence between the system and the bath.\cite{Dijkstra.2010.prl.104.250401} In order to treat the bath quantum mechanically, we employ the hierarchy of equations of motion method.

The hierarchy of equations of motion method was initially developed for a system coupled to an overdamped vibration. This case is now well known.\cite{Tanimura.1989.jpsj.58.101, Tanimura.2006.jpsj.75.082001, Ishizaki.2008.chemphys.347.185, Ishizaki.2009.jcp.130.234111, Strumpfer.2009.jcp.131.225101, Kreisbeck.2011.jctc.7.2166, Kreisbeck.2012.jpcl.3.2828, Xu.2013.jcp.138.024106}
Overdamped vibrations protect long-lived electronic coherence, when the time scale of the vibrational damping is treated properly. \cite{Tanimura.1994.jpsj.63.66, Ishizaki.2009.jcp.130.234111, Ishizaki.2009.pnas.106.17255, Tanimura.2014.jcp.141.044114} Less is known about the situation with a main system coupled to an underdamped vibration,\cite{Tanaka.2009.jpsj.78.073802, Tanaka.2010.jcp.132.214502, Tanimura.2012.jcp.137.22A550, Kreisbeck.2012.jpcl.3.2828} which is our focus here. Underdamped vibrations can dynamically bring charge transfer and exciton states into resonance, leading to rates of irreversible charge transfer that are impossible without vibronic coupling. 

Although it is possible to model a more general vibrational mode by employing the hierarchy in the Wigner picture,\cite{Tanimura.1994.jcp.101.3049, Tanimura.1997.jcp.107.1779, Tanimura.2015.arxiv, Sakurai.2013.jpsj.82.033707, Sakurai.2014.njp.16.015002, Yao.2014.jcp.140.104113} we here use the simpler model of a harmonic potential. The main electronic system of interest, modeled by a Hamiltonian $H_S$ is coupled to a single harmonic mode, which in turn is coupled to a bath of infinitely many harmonic modes. These modes lead to damping of the primary vibrational mode. We will choose parameters in such a way that the damping is in the underdamped regime, so that the vibration vibrates. The spectral density for the coupling of the primary vibration to its environment is chosen to be of Ohmic form. Through a transformation, the model can be transformed into an electronic system coupled to a bath of infinitely many harmonic vibrations, with an altered spectral density.\cite{Tanaka.2009.jpsj.78.073802}

The Hamiltonian of the model is then
\begin{equation}
H = H_S + \sum_\alpha \left( \frac{p_\alpha^2}{2 m_\alpha} + \frac{1}{2} m_\alpha \omega_\alpha^2(x_\alpha - \frac{c_\alpha}{m_\alpha \omega_\alpha^2} V)^2 \right),
\end{equation}
where $\alpha$ indexes the bath modes, $p$, $m$ and $x$ are the momentum, mass and coordinate, respectively, of the bath oscillator, $c$ is the strength of the coupling of the bath oscillator to the system and $V$ is a system operator. $H_S$ is the system Hamiltonian. The last term in the Hamiltonian corrects for the bath-induced renormalization.

Because the model of linear coupling to a harmonic bath corresponds to Gaussian statistics, all information about the system bath coupling is encoded in the spectral density and the temperature of the bath. 
The spectral density of the Brownian oscillator model (underdamped vibration) is
\begin{equation}
J(\omega) = 2 \hbar \lambda \frac{\gamma\omega_0^2 \omega}{(\omega_0^2-\omega^2)^2 + \gamma^2 \omega^2},
\end{equation}
which has a characterictic frequency $\omega_0$, damping rate $\gamma$ and reorganization energy $\lambda$. 

To construct the equations of motion, one needs the quantum correlation function $L(t)$, which can be calculated in the standard way from the spectral density with the equation
\begin{equation}
  L(t) = L_2(t) - i L_1(t) = \frac{1}{\pi} \int_0^\infty \mathrm{d}\omega J(\omega)(\coth \frac{\beta\hbar\omega}{2} \cos \omega t - i \sin \omega t).
\end{equation}

After performing the integration, e.g. by contour integration, on finds the result to be\cite{Tanaka.2009.jpsj.78.073802}
\begin{equation}
L_1(t) = \frac{\hbar\lambda\omega_0^2}{2 i \zeta} { e^{-(\gamma/2 - i \zeta)t} - e^{-(\gamma/2 + i \zeta)t} }
\end{equation}
and
\begin{eqnarray}
L_2(t) &=& \frac{\hbar \lambda\omega_0^2}{2\zeta} e^{-(\gamma/2-i\zeta)t} \coth \frac{\beta\hbar}{2}(\zeta+ i \frac{\gamma}{2}) \\
 &-& \frac{\hbar \lambda\omega_0^2}{2 \zeta} e^{-(\gamma/2 + i \zeta)t} \coth \frac{\beta\hbar}{2}(-\zeta + i \frac{\gamma}{2}) \\
&-& \frac{4 \lambda\gamma\omega_0^2}{\beta} \sum_{k=1}^\infty \frac{\nu_k}{(\omega_0^2+\nu_k^2)^2 - \gamma^2 \nu_k^2} e^{-\nu_k t},
\end{eqnarray}
where $\zeta=\sqrt{\omega_0^2-\gamma^2/4}$ and $\nu_k = 2 \pi k / \beta \hbar$.

Because the correlation function is a sum of exponentials, a hierarchy of equations of motion for the reduced density matrix can be derived in the usual way. If we write the correlation function as
\begin{equation}
  L(t) = \sum_k A_k e^{-\gamma_k t},
\end{equation}
the hierarchy is given by
\begin{eqnarray}
\dot \rho^n(t) &=& -\left( i H_S'^\times + \sum_k n_k \gamma_k \right) \rho^n(t) \nonumber \\
               &+& \sum_k [V, \rho^{n_k^+}] \nonumber \\
               &+& \sum_k n_k \left( A_k V \rho^{n_k^-} + A_k^* \rho^{n_k^-} V_k \right).
\end{eqnarray}
$n_k^+$ ($n_k^-$) refer to an increase (decrease) of the respective index by one. 
The only differences with the original hierarchy for the Drude Lorentz spectral density are the presence of an extra dimension and the different coefficients $A$ and $\gamma$. In order to calculate two-dimensional spectra, a separate hierarchy is used for the three times $t_1$, $t_2$ and $t_3$. At the moment of interaction with the light, all tiers of the hierarchy are multiplied with the dipole operator, which leads to the correct preservation of memory over the external interaction, and the resulting elements are copied from $t_1$ to $t_2$ or from $t_2$ to $t_3$. During $t_1$, only a coherence between the ground state and the excited state is present, during $t_2$ populations and coherences in the one-particle manifold are included, while during $t_3$ again only coherences are considered.

\begin{figure}[t]
\includegraphics[width=8cm]{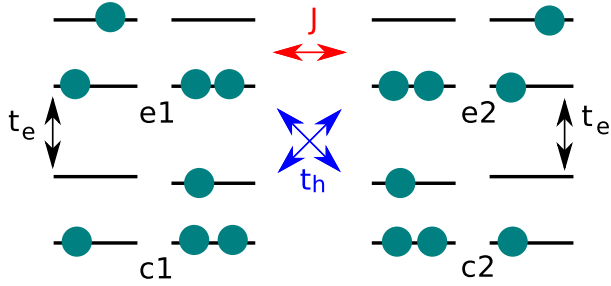}
\caption{\label{fig:cartoon} The states in the system Hamiltonian. Horizontal lines denote the HOMO and LUMO levels of the two molecules and circles show electrons. The states labeled $e_1$ and $e_2$ are exciton basis states, those labeled $c_1$ and $c_2$ are charge transfer states.}
\end{figure}

Now that we established the Hamiltonian for the system coupled to an underdamped vibration, it remains to specify the model Hamiltonian for the electronic system. The simplest possible model system which is directly relevant for the study of photosynthetic complexes, and which contains charge transfer as well as exciton transfer effects, is an electronic dimer.\cite{Hayes.2013.science.430.6139, Halpin.2014.natchem.6.196}
In order to describe a system with electron, hole, and exciton transfer, we take as our system Hamiltonian
\begin{eqnarray}
  H_\mathrm{S} &=& E_\mathrm{CT} (|c1\rangle \langle c1| + |c_2\rangle \langle c_2|) \nonumber \\
  &+& J (|e_1\rangle \langle e_2| + |e_2\rangle \langle e_1|) \nonumber \\
  &+& t_h (|e_2\rangle \langle c_1| + |c_1\rangle \langle e_2|
   + |e_1\rangle \langle c_2| + |c_2 \rangle \langle e_1|) \nonumber \\
  &+& t_e (|e_1\rangle \langle c_1| + |c_1\rangle \langle e_1|
   + |e_2\rangle \langle c_2| + |c_2 \rangle \langle e_2|),         
\end{eqnarray}
where $|e_1\rangle$ and $|e_2\rangle$ are the exciton states, $|c_1\rangle$ and $|c_2\rangle$ are the charge transfer states, $E_\mathrm{CT}$ is the energy of the charge transfer states, $J$ is the excitonic coupling, $t_h$ is the hole transfer and $t_e$ the electron transfer. Note that the Hamiltonian has four excited states. In the calculations reported in the following section, we set $t_h = 0.1$ or 0, $t_e = 0.1$ or 0 and $J = -0.25$ or 0. The energy of the charge transfer states is set to $E_\mathrm{CT} = -0.3$, while the reorganization energy is $\lambda = 2.0$, the damping $\gamma = 0.2$ and the inverse temperature $\beta = 1.5$. All these values are scaled to the vibrational frequency, which is set to $\omega_0=1$. The parameters can easily be rescaled to the values relevant for real systems. Because for a typical photosynthetic system, the largest J-couplings are in the range of 50-150 cm$^{-1}$, the parameters used here correspond to vibrational frequencies of several hundred wavenumbers. Such vibrations are, indeed, ubiquitous in photosynthetic systems. Our model parameters are therefore directly relevant to real systems. Although the reorganization energy chosen here is rather large, such values are expected, for example, for DNA bases.\cite{Dijkstra.2010.njp.12.055005} The states that appear in the system Hamiltonian are shown in Fig.~\ref{fig:cartoon}.

The final model parameter that needs to be specified is the way the vibrational bath interacts with the electronic system. Although our approach can be applied to more general system bath interactions, we here choose the coupling is such a way that the vibrational bath affects the electron transfer. Therefore, the system part of the system bath coupling is 
\begin{equation}
  V = \frac{1}{2} (- |e_1\rangle \langle e_1| + |e_2\rangle \langle e_2|
                   + |c_1\rangle \langle c_1| - |c_2\rangle \langle c_2| ).
\end{equation}
The renormalization term which contains $V^2$ is added to the system Hamiltonian, $H_S' = H_S + \lambda V^2$, with the reorganization energy $\lambda = \int_0^\infty \mathrm{d}\omega J(\omega) / \pi \omega$. This term corrects the bath induces shift in the system parameters.

For the transition dipole operator, which couples the exciton states to the ground state, we have
\begin{equation}
  \mu = \mu_1 (|g\rangle \langle e_1| + |e_1\rangle \langle g|)
      + \mu_2 (|g\rangle \langle e_2| + |e_2\rangle \langle g|)
\end{equation}
In particular, we choose the two transition dipoles to be parallel and of equal length, so that we can ignore the vector nature of the transition dipoles and set $\mu_1 = \mu_2 = 1$. This assumption can easily be relaxed if one is interested in the effect of laser polarization on the two-dimensional spectra. 

\begin{figure}[t]
\includegraphics{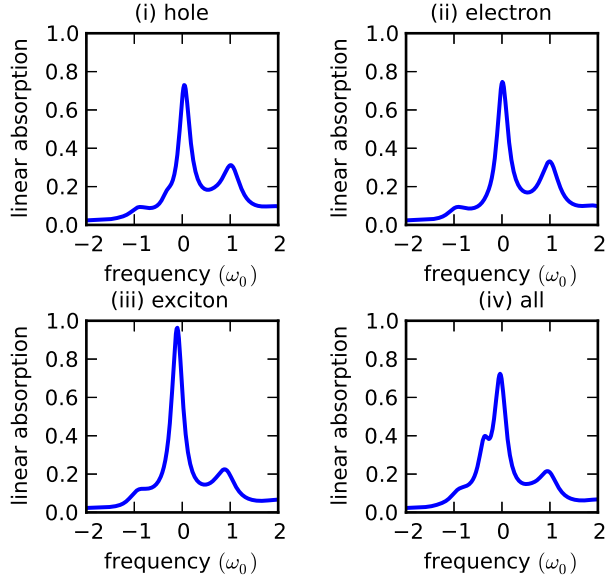}
\caption{\label{fig:speclin} Linear spectrum with (i) hole transfer only, (ii) electron transfer only, (iii) exciton transfer only and (iv) all three modes of transfer together, as indicated above the respective panels.}
\end{figure}

\section{Results and Discussion} \label{sec:results}

\subsection{Linear spectra} 
First, we turn our attention to the linear spectra. Because population dynamics only affects the line shape of the spectrum, this technique is not very sensitive to the details of the coupling of exciton to charge transfer states, i.e. of the charge transfer mechanism. However, linear spectra already contain a wealth of information. The peak positions reveal the energies of underlying states, while their shape reflects the interaction with the environment. Although charge transfer states, which are optically dark, are not directly visible by themselves, they lead to observable effects on the linear spectra because they couple to the bright exciton states. In addition to exciton states, underdamped vibrations have a directly observable effect on the linear spectrum. They lead to vibrational side bands in addition to the main exciton peak, with positions determined by the vibrational frequency and intensities derived from the Huang Rhys factors. What is not known, and is an important goal of our simulations, is the information content of linear spectra in the case where both charge transfer and coupling to underdamped vibrations is present. 

In the linear spectra, shown in figure \ref{fig:speclin} we observe an interesting effect. In order to identify the effect of charge transfer, we compare spectra where (i) only hole transport is present, where (ii) only electron transport is present, where (iii) only exciton transport is present, and where (iv) all mechanisms contribute. At a first glance, we observe that a main peak and vibrational side bands are present in all spectra. The spectra for (i) hole ($t_h = 0.1, t_e = J = 0$) or (ii) electron coupling ($t_e = 0.1, t_h = J = 0$) only look very similar, apart for a weak shoulder at the red side of the main peak in the hole transfer spectrum. When we consider the other two spectra, however, some differences are observed. The spectrum for (iii) exciton transfer ($J=-0.25, t_h = t_e = 0$) only is somewhat larger in intensity and shifted. Also in this case, the vibrational side bands are similar to the hole and electron transfer cases. However, a surprise is present in the case where (iv) all three modes of transport are present ($t_h = t_e = 0.1, J=-0.25$): the shoulder on the red side of the main peak is now very pronounced. This shows that electron, hole and exciton transfer cannot be considered separately, but must all be included in order to reproduce the full result already for the linear spectrum. The observation of an extra peak in the linear spectrum of a charge transfer system coupled to an underdamped vibration is the first main finding of this paper. 

\begin{figure}[t]
\includegraphics{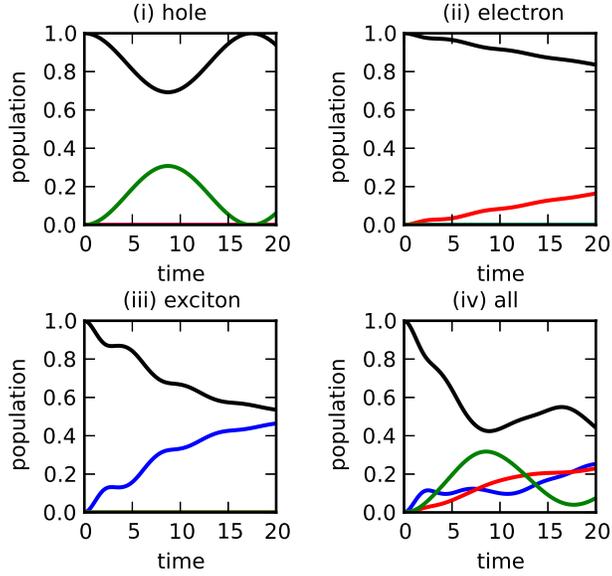}
\caption{\label{fig:dynam} Population dynamics of exciton states (black and blue line) and charge transfer states (red and green line) for (i) hole transfer only, (ii) electron transfer only, (iii) exciton transfer only and (iv) the complete Hamiltonian. The initial condition is all population on one of the exciton states. Time is in units of $1/\omega_0$.}
\end{figure}

\subsection{Dynamics} 
We first look at the dynamics for (i) hole transport only, (ii) electron transport only, (iii) exciton transport only and (iv) the complete picture with all tranfer modes combined. The dynamics can be understood as taking place in a quantum network\cite{Cao.2009.jcpa.113.13825} coupled to an underdamped vibration. In this picture, a clear distinction can be seen between coherent hole transport on the one hand and mostly incoherent electron and exciton transport on the other hand. Figure \ref{fig:dynam} shows the dynamics, which follows intuitive behavior. Vibrational oscillations are weak, although they are visible in the exciton transfer. Electron and hole transport couple the initially populated exciton state to a charge transfer state. Exciton transport depopulates the initially excited state and transfers the population to the other exciton state. The different dynamics can be explained as follows. Hole transport (i) is not affected by the coupling to the vibration, and therefore exhibits coherent Rabi oscillations. Electron transfer (ii), on the other hand, is damped due to the interaction with the vibration, and is incoherent. Exciton transport (iii) is also mostly incoherent, as a result of the effect of the damped vibration. When all three transfer mechanisms are present (iv), a combination of partial coherent oscillations and partial incoherent transport are observed. Because the system is in a population or excited state coherence during the waiting time of a two-dimensional experiment, one would expect to see similar behavior when plotting the waiting time dependence of peaks in the two-dimensional spectrum. Because the dynamics in the four cases investigated here are clearly different, one may hope that two-dimensional spectroscopy can distinguish them. We will see, however, that this is not the case.

\begin{figure}[t]
\includegraphics{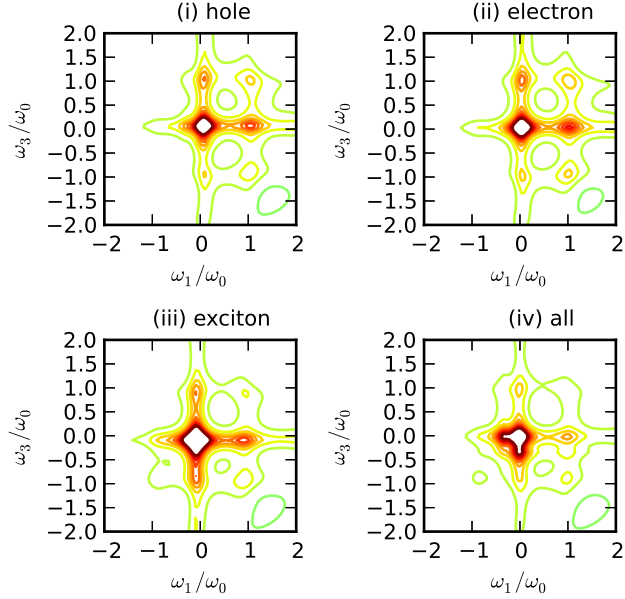}
\caption{\label{fig:2D} Two-dimensional spectrum with (i) hole transfer only, (ii) electron transfer only, (iii) exciton transfer only and (iv) all three modes of transfer together, as indicated above the respective panels. The waiting time is 10.0 / $\omega_0$.}
\end{figure}

\subsection{Two-dimensional spectra} 
In figure \ref{fig:2D} we present calculated two-dimensional correlation spectra. Because the effect of excited state absorption complicates the discussion, and leads to peaks of opposite sign which can usually be separated from bleaching and stimulated emission peaks, we focus here on the latter two contributions.\cite{Huynh.2013.jcp.139.104103} The inclusion of excited state absorption is left as a possible extension in future work.

The first obvious effect in the two-dimensional spectra is the presence of the extra feature due to the interplay of charge transfer with vibrations observed already in the linear spectrum. Here, as shown in figure \ref{fig:2D} (d) we observe the same peak appear as a cross peak with the main exciton absorption feature, which partially overlaps with this main diagonal peak. In addition, we observe that the vibrational cross peak in the spectrum with all interactions present is weaker than when only electron, hole, or exciton transfer is present. Similar to our analysis of the linear spectrum, we conclude that cross peaks in two-dimensional spectra can not be interpret as arising from charge transfer, exciton transfer or vibrations alone, but that they are the result of a complex interplay of all these ingredients.

\begin{figure}[t]
\includegraphics{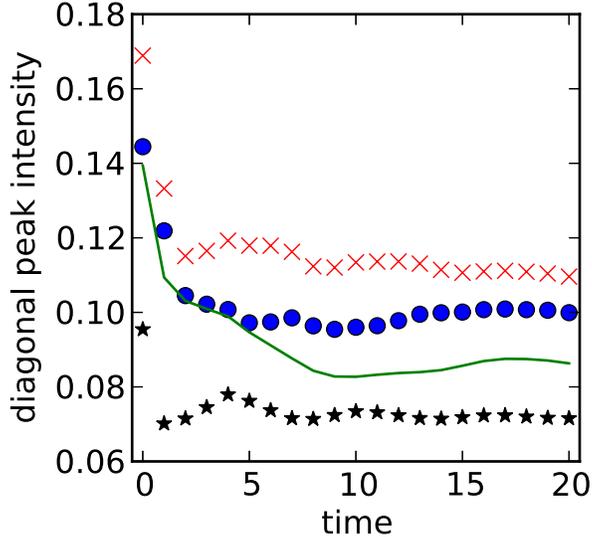}
\caption{\label{fig:diag} Main diagonal peak intensity as a function of waiting time for (blue circles) hole transport only, (red crosses) electron transport only, (black stars) exciton transport only, and (green line) the full Hamiltonian. Time is in units of $1/\omega_0$.}
\end{figure}

\begin{figure}[t]
\includegraphics{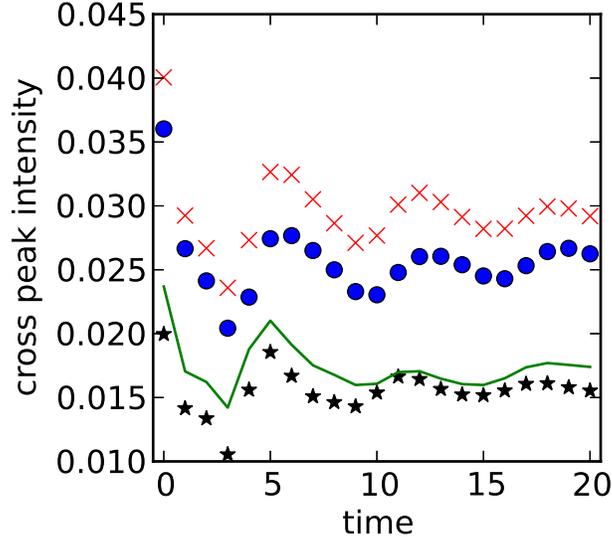}
\caption{\label{fig:cross} Vibrational cross peak intensity as a function of waiting time for (blue circles) hole transport only, (red crosses) electron transport only, (black stars) exciton transport only, and (green line) the full Hamiltonian. Time is in units of $1/\omega_0$.}
\end{figure}

\subsection{Waiting time dependence} 
In order to analyze the two-dimensional spectrum further, we plot the intensity of the diagonal peak ($(\omega_1, \omega_3) = (0,0)$) and one of the vibrational cross peaks ($(\omega_1, \omega_3) = (1,0)$) as a function of the waiting time, shown in figures \ref{fig:diag} and \ref{fig:cross}. Because populations of exciton and charge transfer states (or, indeed, coherent superpositions of these) are present during the waiting time, one would expect the dynamics of these states to be reflected directly in these time traces. We would expect to see clear coherent oscillations in the case of hole transport, corresponding to the oscillations found in the dynamics, while such oscillations should be absent in the mostly incoherent electron and exciton transfer. However, we hardly see this effect in the calculated two-dimensional spectra. In fact, all time traces show similar oscillations. We conclude that, in contrast to the dynamics, oscillations in the spectra are dominated by vibrational coherence. This is the second main finding of this paper. The fact that the electron and hole dynamics are invisible can be attributed to the fact that the charge transfer states don't couple directly to the light, but interact only via the exciton states.

\subsection{Fifth order response}

\begin{figure}[t]
\includegraphics{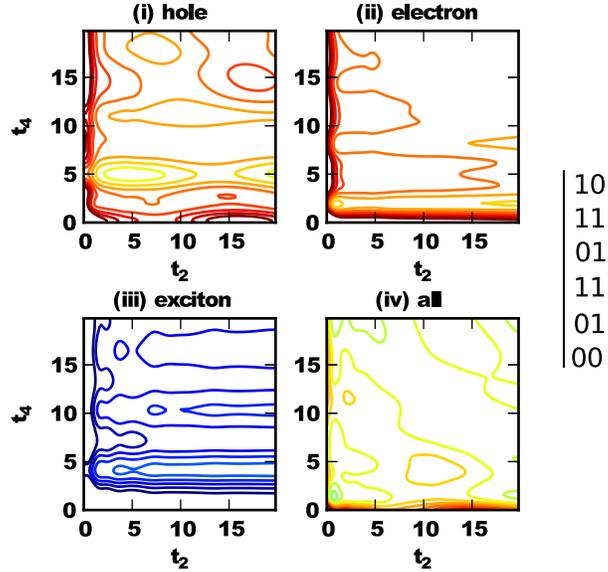}
\caption{\label{fig:fifthorder} Fifth-order response function as a function of the population times $t_2$ and $t_4$ for (i) hole transport only, (ii) electron transport only, (iii) exciton transport only and (iv) the full Hamiltonian. Note the opposite sign of the response in the case of exciton dynamics. Contours in red and blue represent positive and negative values, respectively. Coherence times $t_1$ and $t_5$ are set to zero, while $t_3=20$. The double-sided Feynman diagram corresponding to the response function calculated here is shown to the right.}
\end{figure}

We continued our search for experimentally observable signatures of the charge transfer process by considering the fifth-order nonlinear response.\cite{Zhang.2012.optlett.37.5058, Zhang.2013.jpcb.117.15369} This order is known to be more sensitive to anharmonicity in certain cases.\cite{Tanimura.1997.jcp.106.2078} The fifth-order response functions contain three coherence times and two population times. Because correlated population dynamics during two time intervals is present, one might hope to see more details of the coupling to charge transfer states. 

In figure \ref{fig:fifthorder} we plot a fifth-order response function in the $k_s = -k_1+k_2-k_3+k_4+k_5$ phase matched direction. This spectroscopy can be called "2D population spectroscopy", because it exhibits the correlations between population dynamics during two separate time intervals. The difference between the different cases is clearly visible, in fact, the response with only exciton transport present has a different sign. Although we calculated only one contribution to the complete fifth-order signal here this finding indicates that the fifth order response is very sensitive to the charge transfer process. This is the third main result of this paper.

Key to this analysis is the fact that changes in population, as observed in figure \ref{fig:dynam} are reflected in the 2D population spectrum. Because of correlations building up during the coherence time $t_3$, the signal is not symmetric along the $t_2=t_4$ line. We note that it is important to set $t_3$ not equal to zero for the observation of the charge transfer dynamics. While not studied in detail here, it was found in previous work that coherent dynamics during the $t_5$ period is also important for the signal.\cite{Tanimura.1997.jcp.106.2078} By changing $t_5$, it is possible to tune the interference of signals arising from excited state or ground state dynamics to enhance or surpress the oscillatory motions of specific physical processes. Although we fixed $t_1=t_5=0$ here, we can utilize a representation similar to the 2D frequency domain spectrum shown in figure \ref{fig:2D} to further visualize the fifth-order response in order to analyze the coherence and population dynamics of each cross peak separately. We leave a further investigation of the fifth-order signal as an interesting direction for future work.

\section{Conclusions} \label{sec:concl}

In summary, we have studied a system in which exciton and charge transfer exist together with coupling to an underdamped vibration. 
This work shows that incorporating electron and hole transfer into an excitonic model can lead to additional peaks in optical spectra. This peak, which is present in the linear spectrum as well as as a cross peak in the two-dimensional spectrum, can not be explained by exciton or charge transfer alone, but is the result of a complex interplay of both these ingredients with vibrational coupling. This finding shows that one has to be careful in the assignment of peaks in experimental spectra to excitonic or vibrational peaks alone, while, in fact, charge transfer may also be an important ingredient. 

Oscillations as a function of the waiting time in the two-dimensional spectrum do not reflect the population dynamics and are attributed to vibrations, even though the effect of the vibration on the dynamics is weak. Although, therefore, extracting information about the charge transfer process from two-dimensional optical spectroscopy is difficult, we find that the fifth-order reponse, in the form of 2D population spectroscopy, is particularly sensitive to this process.

Our results should help the interpretation of spectra of photosynthetic light harvesting systems,\cite{Engel.2007.nature.446.782, Collini.2010.nature.463.644} conjugated polymers,\cite{Collini.2009.science.323.369} model dimers\cite{Hayes.2013.science.430.6139, Halpin.2014.natchem.6.196} and DNA.\cite{Dijkstra.2010.njp.12.055005} Further extension of the model presented here to calculate the specific properties of these systems in more detail is relatively straightforward and is left as a direction for future work. We also hope that our results will stimulate experimental and theoretical work on fifth-order nonlinear processes as a probe of charge transfer. An investigation of the full fifth order response for a model system is under way.

\section*{Acknowledgements}
This research was supported by a Marie Curie International Incoming Fellowship within the 7th European Community Framework Programme (grant no. 627864).

\bibliography{references}

\end{document}